%% file: main.tex
\documentclass[journal]{IEEEtran}

\usepackage{cite}

\ifCLASSINFOpdf
   \usepackage[pdftex]{graphicx}
   \graphicspath{{../pdf/}{../jpeg/}}
   \DeclareGraphicsExtensions{.pdf,.jpeg,.png}
\else
\fi
\UseRawInputEncoding
\usepackage{amsmath}
\usepackage{array}
\usepackage{standalone}
\usepackage{etoolbox}
\newtoggle{arXiv}
\newtoggle{bibrun}
\usepackage{newtxtext, newtxmath}
\usepackage{graphicx}
\usepackage{color, calc}
\usepackage{array, booktabs}
\usepackage{tikz, tikz-3dplot}
\usetikzlibrary{arrows, arrows.meta, calc, positioning}
\usetikzlibrary{decorations.pathmorphing}	
\tikzset{>=Stealth}
\usepackage{siunitx, physics}

\usepackage{enumitem}
\setlist[description]{labelindent=0pt, leftmargin=\parindent, font=\normalfont\itshape}
\usepackage{url}
\usepackage{subfigure}
\usepackage{textcomp}
\usepackage{eurosym}
\usepackage{xfrac}
\usepackage{pgfplots}
\usepackage{stmaryrd}
\pgfplotsset{compat=1.17}
\usepackage{makecell}

\newcommand{\m}{MaRCoS}

\begin{document}

\title{Benchmarking the performance of a low-cost Magnetic Resonance Control System at multiple sites in the open \m{} community}

\author{\IEEEauthorblockN{
		Teresa Guallart-Naval\IEEEauthorrefmark{2}$^,$\IEEEauthorrefmark{3},
		Thomas O'Reilly\IEEEauthorrefmark{4},
		Jos\'e M. Algar\'in\IEEEauthorrefmark{2},
		Ruben Pellicer-Guridi\IEEEauthorrefmark{5},
		Yolanda Vives-Gilabert\IEEEauthorrefmark{6},
		Lincoln Craven-Brightman\IEEEauthorrefmark{7},
		Vlad Negnevitsky\IEEEauthorrefmark{8},
		Benjamin Menk\"uc\IEEEauthorrefmark{9},
		Fernando Galve\IEEEauthorrefmark{2},
		Jason P. Stockmann\IEEEauthorrefmark{7},
		Andrew Webb\IEEEauthorrefmark{4},
		and	Joseba Alonso\IEEEauthorrefmark{2}}
	
	\IEEEauthorblockA{\IEEEauthorrefmark{2}MRILab, Institute for Molecular Imaging and Instrumentation (i3M), Spanish National Research Council (CSIC) and Universitat Polit\`ecnica de Val\`encia (UPV), Valencia, Spain}\\
	\IEEEauthorblockA{\IEEEauthorrefmark{3}Tesoro Imaging S.L., Valencia, Spain}\\
	\IEEEauthorblockA{\IEEEauthorrefmark{4}Department of Radiology, Leiden University Medical Center, Leiden, Netherlands}\\
	\IEEEauthorblockA{\IEEEauthorrefmark{5}Asociaci\'on de investigaci\'on MPC, Donostia-San Sebasti\'an, Spain}\\
	\IEEEauthorblockA{\IEEEauthorrefmark{6}Intelligent Data Analysis Laboratory, Department of Electronic Engineering, Universitat de Val\`encia, Valencia, Spain}\\
	\IEEEauthorblockA{\IEEEauthorrefmark{7}Massachusetts General Hospital, A. A. Martinos Center for Biomedical Imaging, Charlestown, MA, United States}\\
	\IEEEauthorblockA{\IEEEauthorrefmark{8}Oxford Ionics Ltd, Oxford, United Kingdom}\\
	\IEEEauthorblockA{\IEEEauthorrefmark{9}University of Applied Sciences and Arts Dortmund, Dortmund, Germany}\\
	
\thanks{Corresponding author: J. Alonso (joseba.alonso@i3m.upv.es).}}

\markboth{Journal of \LaTeX\ Class Files,\,Vol.\,X, No.\,X, MARCH\,2022}%
{Shell \MakeLowercase{\textit{et al.}}: Bare Demo of IEEEtran.cls for IEEE Journals}

\maketitle

\begin{abstract}
\newline
Purpose: {\normalfont To describe the current properties and capabilities of an open-source hardware and software package that is being developed by many sites internationally with the aim of providing an inexpensive yet flexible platform for low-cost MRI.}
\\
Methods: {\normalfont This paper describes three different setups from 50 to 360\,mT in different settings, all of which used the MaRCoS console for acquiring data, and different types of software interfaces (custom-built GUI or PulSeq overlay) to acquire the data.}
 \\
Results: {\normalfont Images are presented from both phantoms and \emph{in vivo} from healthy volunteers to demonstrate the image quality that can be obtained from the MaRCoS hardware/software interfaced to different low-field magnets.}
 \\
Conclusions: {\normalfont The results presented here show that a number of different sequences commonly used in the clinic can be programmed into an open-source system relatively quickly and easily, and can produce good quality images even at this early stage of development. Both the hardware and software will continue to develop, and it is an aim of this paper to encourage other groups to join this international consortium.}
\end{abstract}


 \ifCLASSOPTIONpeerreview
 \begin{center} \bfseries EDICS Category: 3-BBND \end{center}
 \fi
%
\IEEEpeerreviewmaketitle


\section{Introduction}

\IEEEPARstart{M}{aRCoS} (Magnetic Resonance Control System) is a low-cost, high-performance console developed to fulfill the requirements of a rapidly expanding low-field Magnetic Resonance Imaging (LF-MRI) community \cite{Negnevitsky2022,Negnevitsky2021,Craven-Brightman2021}. LF-MRI is developing as a customizable and affordable complement to standard high-field MRI ($> 1$\,T), which is an expensive medical imaging modality in terms of purchase cost, maintenance, siting and training, and consequently is concentrated in large hospitals in the economically developed world \cite{Sarracanie2015,Marques2019,Sarracanie2020,Wald2020,Bhat2021}. In the last few years, LF-MRI has demonstrated its value for point-of-care imaging \cite{McDaniel2019,Nakagomi2019,Cooley2020,OReilly2020,Sheth2021,Mazurek2021,Liu2021}, home healthcare \cite{Guallart-Naval2022}, quantitative MRI and fingerprinting \cite{OReilly2021,Sarracanie2021}, hard-tissue imaging \cite{Algarin2020,Gonzalez2021,Borreguero2022}, artifact-free imaging of metallic implants \cite{Guallart-Naval2022,VanSpeybroeck2021}, as well as educational purposes \cite{Cooley2020b}, among others. These achievements are enabled by a new generation of scanners combining refined hardware engineering with powerful computational algorithms, including machine learning architectures. Improvements are being integrated in all engineering and imaging stages: during the scanner design process \cite{OReilly2019,DeVos2020,Wenzel2021}, for pulse sequence design \cite{Sarracanie2021}, during signal acquisition \cite{Liu2021}, for image reconstruction \cite{Algarin2020,Koonjoo2021}, and for data analysis and image post-processing \cite{Arnold2022}.

The LF-MRI community is integrated by numerous and diverse research groups and spin-off companies, each targeting specific applications. Consequently, the developed LF-MRI scanners are often special-purpose and custom-made. Different groups have historically pursued different solutions as regards the console (or electronic control system) which compiles the pulse sequences, executes them, digitizes the detected signals and frequently also processes and displays the reconstructed MR images. Some employ commercially available systems such as from Pure Devices GmbH (Rimpar, Germany), Magritek Ltd (Wellington, New Zealand) or Niumag Corporation (Suzhou, China). Despite being much less expensive than clinical consoles, these nevertheless constitute a substantial fraction of the total cost of an LF-MRI system, and are often run with proprietary and non-interchangeable software/hardware. These drawbacks limit prototype production by a wider community and hinder new developments. To overcome these limitations, several home-spun designs have emerged during the last decade, mainly based on Field Programmable Gate Arrays (FPGA)  \cite{Stang2012,Tang2015,Hasselwander2016,Anand2018,OCRA,OpenCoreNMR,LimeSDR,Ang2017}. The \m{} initiative is an attempt to centralize these efforts to become a versatile, low-cost and open-source solution fostering the development of present and future LF-MRI technologies. The \m{} community is therefore open and has the ambition of expanding the range of tools this platform provides as new needs arise. We now have a first functional version of \m{} and a Python-based Graphical User Interface (GUI), both publicly available in open repositories \cite{MaRCoS,MaRCoSGUI}.

In this paper, we benchmark the performance of \m{} in three different LF-MRI systems: a 360\,mT educational tabletop scanner at Massachusetts General Hospital (MGH) in Boston, USA; a 50\,mT low-cost human Halbach system for brain and extremity imaging at Leiden University Medical Center (LUMC) in Leiden, the Netherlands; and a similar Halbach system (70\,mT) at the Institute for Molecular Imaging and Instrumentation (i3M) in Valencia, Spain. After a brief introduction to the \m{} architecture and the individual systems in Sec.\,\ref{sec:methods}, we demonstrate \m{}' versatility in terms of imaging capabilities with a variety of pulse sequences (Sec.\,\ref{sec:phantoms}). To this end, we have programmed and run spin echo (SE), rapid imaging with refocused echoes (RARE), gradient echo (GRE), short tau inversion recovery (STIR), and non-Cartesian pulse sequences to image phantoms with different contrasts, resolutions and reconstruction techniques across all three sites. In addition, we present in Sec.\,\ref{sec:invivo} the first \emph{in vivo} images taken with \m{} on the i3M system. These images demonstrate the value of \m{} for potential screening applications, which we discuss in Sec.\,\ref{sec:concl}.

\section{Methods}
\label{sec:methods}

\subsection{\m{}}
\begin{figure}
	\centering
	\includegraphics[width=1\columnwidth]{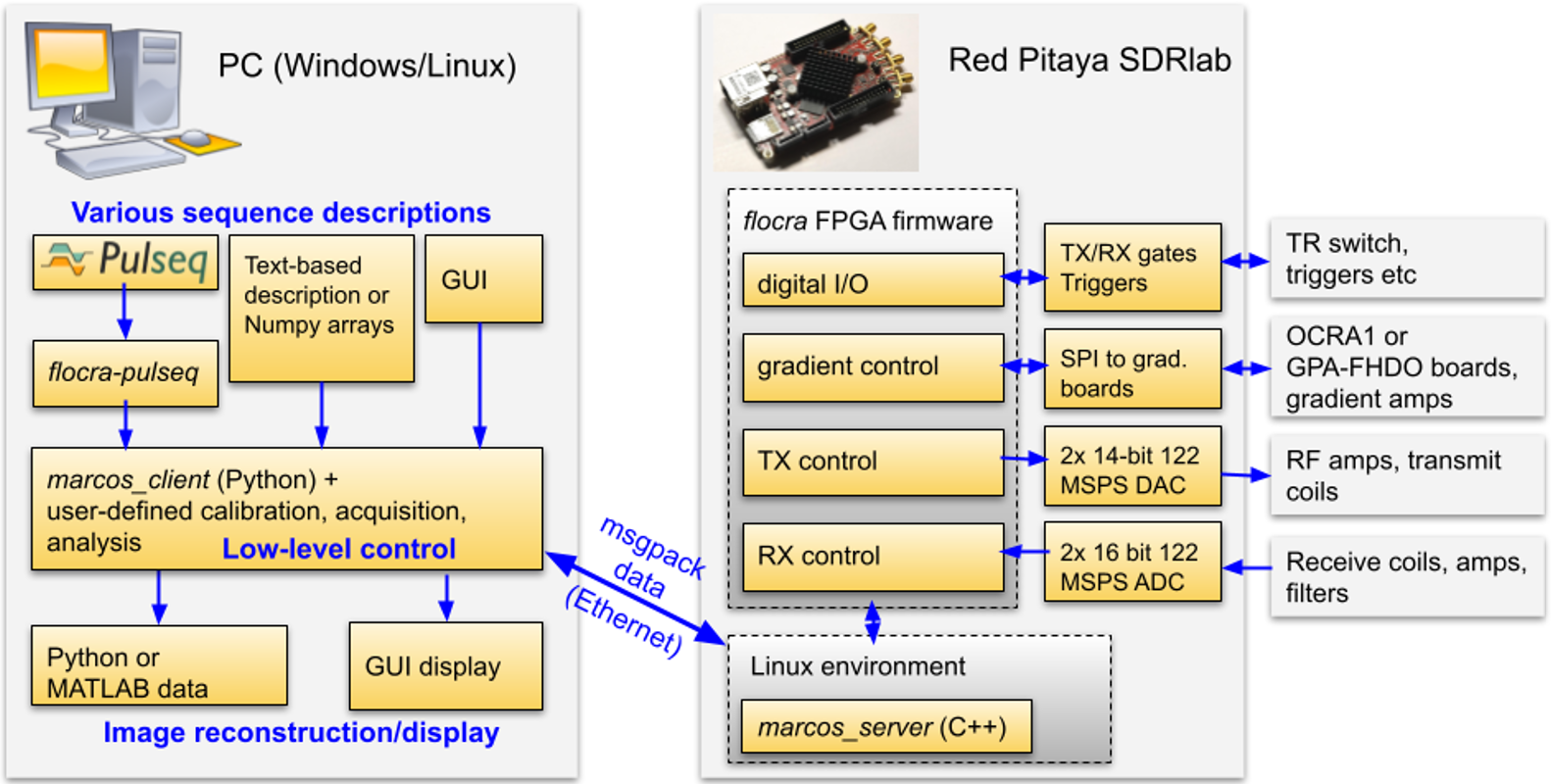}
	\caption{\m{} system architecture, showing the main components in the software and hardware stacks. Figure taken from Ref.\,\cite{Negnevitsky2022}.}
	\label{fig:marcos}
\end{figure}

All images in this work have been acquired with \m{}. Among the various control systems developed by the LF-MRI community, we chose to create \m{} based on OCRA (Open-source Console for Real-time Acquisition, \cite{Anand2018,OCRA}) because it included: i) mature and affordable hardware available off-the-shelf with ample expansion possibilities, and ii) validated open-source firmware and software designs.

The \m{} system architecture is based on a Red Pitaya SDRlab (Fig.\,\ref{fig:marcos}), following the OCRA design. The Red Pitaya includes an ARM processor, a Xylinx Zynq FPGA, two fast analog inputs, to fast analog outputs, and several digital input and output ports. The FPGA firmware (\emph{flocra}) digitally controls the peripherals, such as radio-frequency (RF) transmit/receive chains and gradient coils used in MRI. The ARM processor runs the \m{} server, which is the intermediate communication layer between the control computer and the FPGA. The control computer is equipped with a Python-based GUI where pulse sequences can be programmed and executed. This GUI translates the sequences and provides all parameters to the \m{} client (also written in Python), which in turn communicates with the client in the Red Pitaya. Once a sequence repetition is executed, the MR signal is recorded and the data is pre-processed by flocra and sent to the client. When the sequence is complete, the data can be processed by user-defined algorithms (e.g. image reconstruction, segmentation, filtering, $k$-space correction, etc.) and displayed on the GUI. Optionally, sequences can be also programmed outside the GUI environment. At the moment, there are two further possibilities: to write them as text or Numpy arrays, or in the PulSeq open-source framework \cite{Layton2017}. The latter is a vendor-agnostic environment that enables sequence programming for a growing range of commercial MRI systems. For more detailed information on \m{} we refer to \cite{Negnevitsky2022}. 

\subsection{LF-MRI systems}
\subsubsection{MGH}
\begin{figure}
	\centering
	\includegraphics[width=1.\columnwidth]{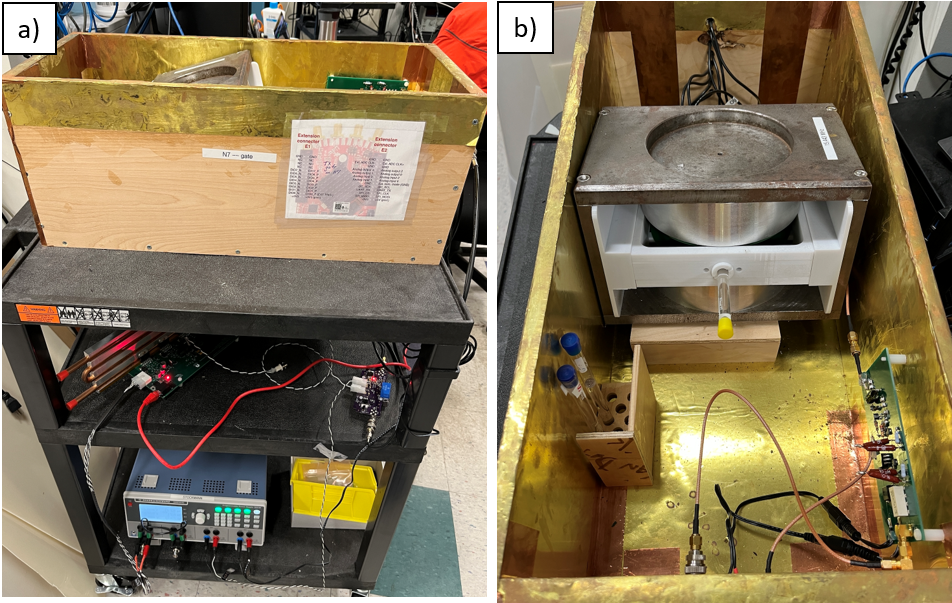}
	\caption{Fully open-source tabletop MRI scanner for education and research at MGH. a) Complete setup including 0.36\,T dipole magnet, FHDO gradient power amplifier board, Red Pitaya-based console, and home-built RF hardware. b) Inner view with phantom inside the scanner.}
	\label{fig:setup_mgh}
\end{figure}

The MaRCoS console was paired with a 0.36\,T tabletop MRI scanner made using open-source hardware components. The tabletop scanner has a 1\,cm field of view (FOV) and was originally developed as an educational tool for university courses at Massachusetts Institute of Technology. Co-planar shielded gradient coils driven by a low-cost, low-voltage gradient power amplifier (GPA-FHDO) \cite{GPAFHDO} are nested between two 6" diameter permanent magnets set up in a dipole configuration. The RF system consists of a small solenoid coil connected to a low-cost TR-switch, preamplifier, and 1\,W RFPA.  The hardware suite was summarized in Cooley \textit{et al.} \cite{Cooley2020b}. The setup was previously paired with a MEDUSA console \cite{Stang2012} and later adapted for use with OCRA. With Red Pitaya-based consoles, the total system cost is below 10\,kUSD. Pulse sequences are written in the PulSeq Matlab environment and then imported into the MaRCoS framework using custom interpreter software (see below).

\subsubsection{LUMC}
\begin{figure}
	\centering
	\includegraphics[width=1.\columnwidth]{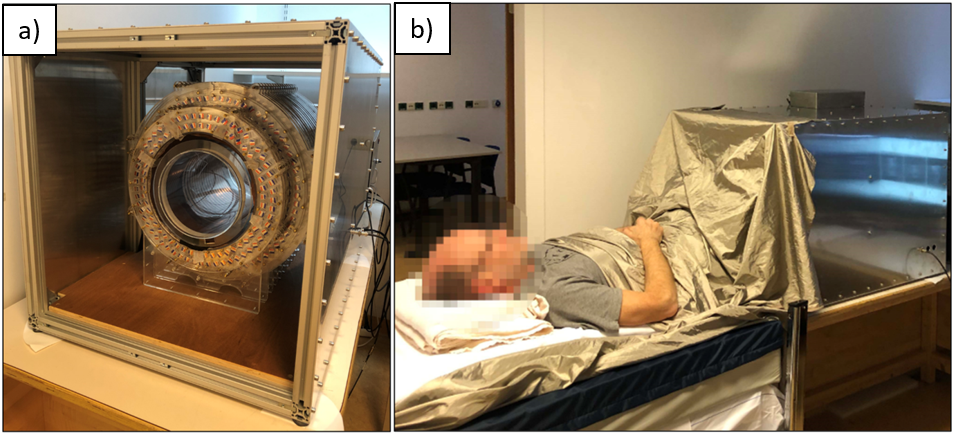}
	\caption{50\,mT system at LUMC. a) Photograph of the scanner inside the Faraday cage. b) Photograph during an \emph{in vivo} acquisition.}
	\label{fig:setup_lumc}
\end{figure}

The custom-built 50\,mT (2.15\,MHz) Halbach-based MRI scanner (Fig.\,\ref{fig:setup_lumc}) is constructed using 2,948 12\,mm cuboid N48 neodymium iron boron magnets arranged in a cylindrical Halbach configuration. The magnet is 50.6\,cm long and has a 27\,cm diameter bore. The magnetic field homogeneity is optimized over a 20\,cm diameter spherical volume (DSV) placed at the centre of the magnet. A set of three linear gradient coils were constructed using a target field method, adapted for the transverse $B_0$ orientation intrinsic to cylindrical Halbach arrays. A custom built 1\,kW RF amplifier with 56\,dB gain is used to amplify the RF pulses. The gradient waveforms are amplified using a custom-built three-axis current-controlled gradient amplifier, powered using two Delta Elektronika SM 18-50  DC power supplies (Zierikzee, the Netherlands). The entire setup is placed inside a Faraday cage constructed from aluminum extrusion and 2\,mm thick aluminum plates. An RF shield is placed inside the inner surface of the bore. During in-vivo experiments the body extends out of the Faraday and couples significant amounts of electromagnetic interference (EMI) into the RF coil. In order to reduce this EMI the body is placed under a conductive cloth (4711 series, Holland Shielding systems BV, Dordrecht, the Netherlands). Further details on the system can be found in Refs.\,\cite{OReilly2019,OReilly2020}.

\subsubsection{i3M}
\begin{figure}
	\centering
	\includegraphics[width=1.\columnwidth]{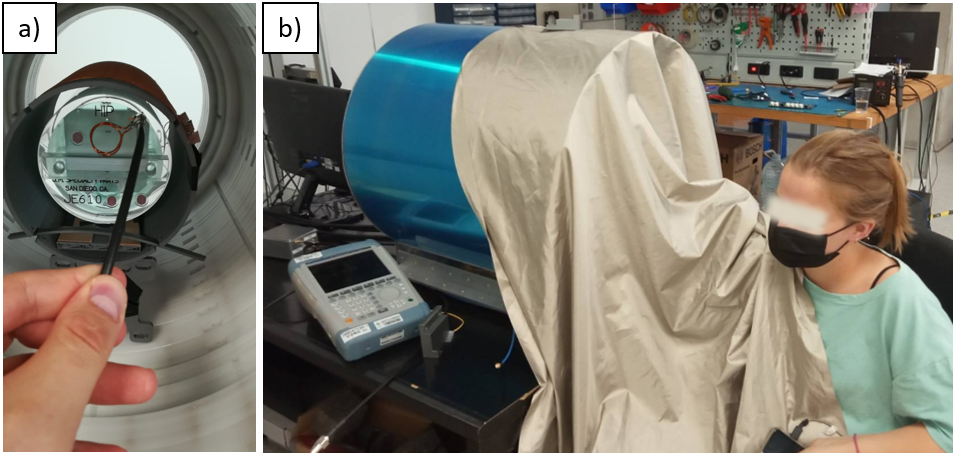}
	\caption{70\,mT system at i3M. a) Photograph of the inside and a phantom used by the American College of Radiation for their MRI accreditation program \cite{Sipila2013}. b) Photograph of the outside and a volunteer during an \emph{in vivo} acquisition.}
	\label{fig:setup_mrilab}
\end{figure}

The \m{}-powered, portable system at i3M is shown in Fig.\,\ref{fig:setup_mrilab} and described in Ref.\,\cite{Guallart-Naval2022}. The scanner design is largely based on the LUMC system described above, with an extra layer of magnets to reach $B_0\approx 72$\,mT. The field homogeneity is shimmed from 15,700 down to 3,100\,ppm (parts per million) over a DSV of 20\,cm in diameter. This is achieved with $\approx1,100$ smaller permanent magnets placed in rings inside and concentric to the main magnet, following an optimized result obtained with a non-linear integer minimization method. The diameter and length of the whole system are around 53 and 51\,cm respectively, and it weighs approximately 200\,kg. The gradient coils are wound on and glued to twelve 3D-printed Nylon molds, and the whole assembly is supported by a methacrylate cylinder. Gradient waveforms are generated with either an OCRA1 board \cite{OCRA1} or the open-source GPA-FHDO board \cite{GPAFHDO} connected to the Red Pitaya via Serial Peripheral Interface (SPI), and amplified by AE Techron 7224 power amplifiers (Indiana, USA). The single transmit/receive (Tx/Rx) RF antenna is a solenoid coil tuned and impedance-matched to the proton Larmor frequency (3.076\,MHz). The RF coil holder was printed in polylactic acid (PLA), and the wire was fixed with cyanoacrylate adhesive. The coil is inside a grounded faraday cage for noise immunity and to prevent interference between the gradients and the RF system, and again a conductive cloth is used for \emph{in vivo} imaging. The RF low-noise and power amplifiers, as well as the Tx/Rx switch, were purchased from Barthel HF-Technik GmbH (Aachen, Germany). 

\subsection{Pulse sequences and image reconstruction}
To give an idea of the potential applications of \m{} we have demonstrated its functionality in the aforementioned MRI scanners. To this end, we show numerous images in Sec.\,\ref{sec:results} which benchmark the performance of \m{} in terms of programming flexibility and its capability to execute multiple pulse sequences and to reconstruct and process the acquired data. All sequence parameters are compiled in Table\,\ref{tab:seq_params} for readability.

\begin{figure*}
	\centering
	\includegraphics[width=2\columnwidth]{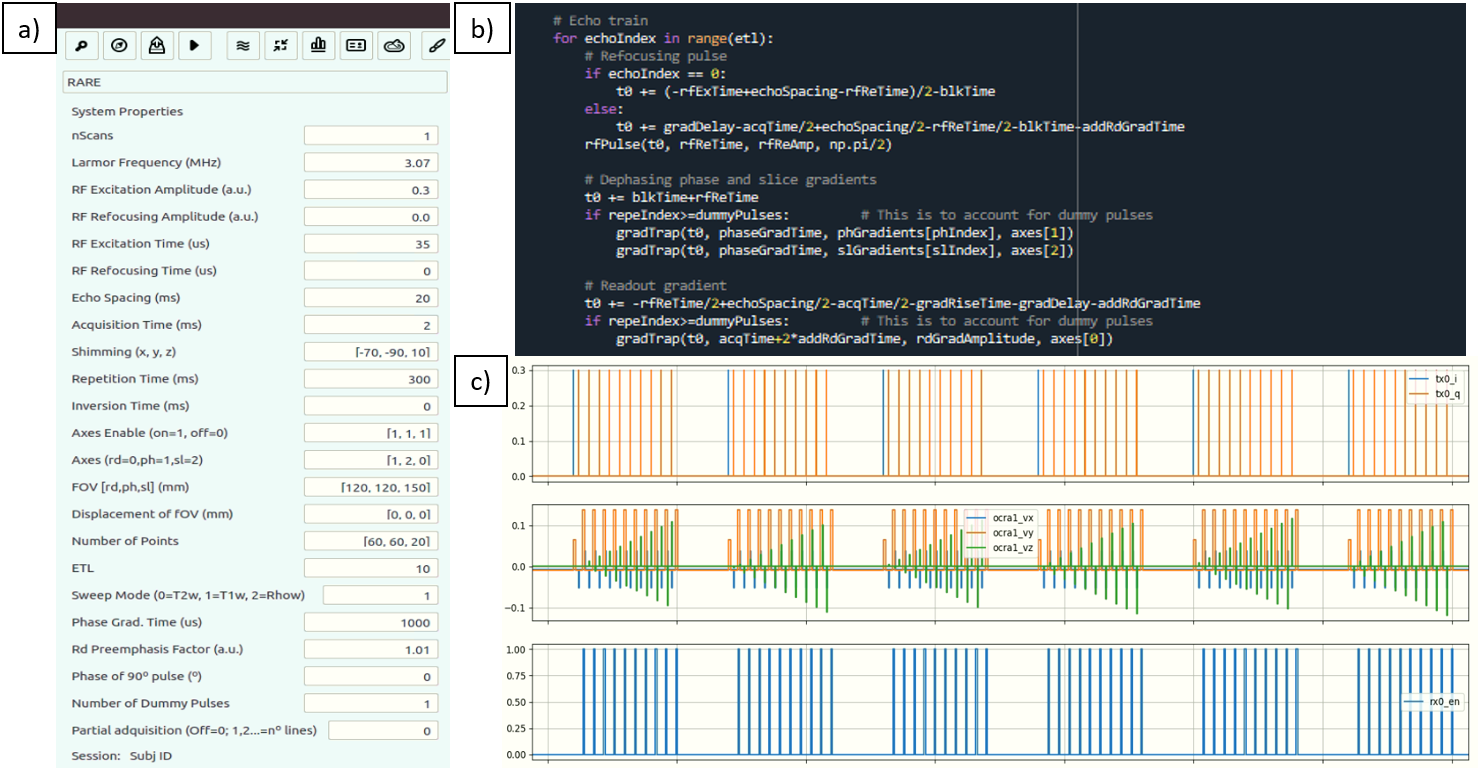}
	\caption{Programming and execution of a 3D-RARE pulse sequence in \m{}: a) GUI window for definition of sequence parameters and execution; b) snippet of Python code; and c) \m{} viewer.}
	\label{fig:seqs}
\end{figure*}

In terms of pulse sequences, in Sec.\,\ref{sec:results} we show images of phantoms taken with SE, RARE, STIR, GRE and radial (non-Cartesian) acquisitions. Figure~\ref{fig:seqs} shows an example of sequence programming (3D-RARE) with the \m{} GUI and a graphical representation of the programmed sequence in the integrated viewer. Furthermore, we have demonstrated low-field images with different contrast mechanisms ($\rho$, $T_1$ and $T_2$) and tested two image reconstruction modules programmed in Python and integrated into the GUI: a standard Fourier Transform (FT) based method; and Non-Uniform Fast Fourier Transform (NUFFT), which can be used for non-Cartesian reconstruction \cite{Greengard2004}.

\begin{table*}
	\caption{Image acquisition parameters. ETL stands for Echo Train Length, and ESP for Echo Spacing. TE in RARE sequences corresponds to the effective echo time.}
	\centering
	\begin{tabular}{c c c c c c c c c c c}
		\hline
		Figure & \makecell{Sample / \\Scanner} & Sequence & \thead{FOV \\ (mm$^3$)} & \# pixels & \thead{BW \\ (kHz)} & \thead{TR \\(ms)} & \thead{TE \\(ms)} & \thead{Other \\params.} & Avgs. & \thead{Scan time \\(min)} \\
		\hline
		
		\hline
		\ref{fig:phantom_mrilab}a) & \makecell{Phantom / \\i3M} & 3D-SE & $130\times130\times130$ & $60\times60\times30$ & 30 & 500 & 20 &  & 1 & 15 \\
        \hline
        
        \ref{fig:phantom_mrilab}b) & \makecell{Phantom / \\i3M} & 3D-RARE & $130\times130\times130$ & $60\times60\times30$ & 30 & 500 & 20 & \makecell{Center-out \\ ETL\,=\,10 \\ ESP\,=\,20\,ms} & 1 & 1.5 \\
        \hline
        
        \ref{fig:phantom_mrilab}c) & \makecell{Phantom / \\i3M} & 3D-STIR & $130\times130\times130$ & $60\times60\times30$ & 30 & 500 & 20 & TI\,=\,20\,ms & 4 & 60 \\
        \hline
        
        \ref{fig:phantom_mrilab}d) & \makecell{Phantom / \\i3M} & 3D-GRE & $130\times130\times130$ & $60\times60\times30$ & 60 & 10 & 2 & Flip angle\,=\,15\,deg & 1 & 9\\
        \hline
        
        \ref{fig:phantom_mrilab}e) \& f) & \makecell{Phantom / \\i3M} & \makecell{3D-SE \\ (radial)} & $130\times130\times130$ & $60\times60\times30$ & 30 & 100 & 20 & \makecell{Radial in $y$, $z$ \\ Equispaced in $x$ \\ Spokes per slice\,=\,95} & 1 & 4.8\\
        \hline

        \hline  
        \ref{fig:phantom_lumc} & \makecell{Phantom / \\LUMC} & 3D-SE & $150\times150\times150$ & $100\times100\times20$ & 20 & 1000 & 30 &  & 1 & 33.3\\
        \hline
        
        \hline  
        \ref{fig:phantom_mgh}) & \makecell{Phantom / \\MGH} & 2D-RARE & $15\times15$ & $128\times128$ & 30 & 5000 & 53 & \makecell{Axial \\ ETL\,=\,32 \\ ESP\,=\,9.5\,ms} & 4 & 1\\
        \hline

        \hline      
        \ref{fig:invivo_mrilab}a) \& e) & \makecell{Elbow / \\i3M} & 3D-RARE & $120\times120\times120$ & $120\times120\times25$ & 30 & 400 & 20 & \makecell{Center-out \\ ETL\,=\,10 \\ ESP\,=\,20\,ms} & 6 & 12 \\
        \hline
       
        \ref{fig:invivo_mrilab}b) \& f) & \makecell{Forearm / \\i3M} & 3D-RARE & $120\times120\times100$ & $120\times120\times25$ & 30 & 400 & 20 & \makecell{Center-out \\ ETL\,=\,10 \\ ESP\,=\,20\,ms} & 6 & 12 \\
        \hline
        
        \ref{fig:invivo_mrilab}c) \& g) & \makecell{Wrist / \\i3M} & 3D-RARE & $200\times160\times80$ & $120\times120\times20$ & 30 & 400 & 20 & \makecell{Center-out \\ ETL\,=\,10 \\ ESP\,=\,20\,ms} & 7 & 11.2 \\
        \hline
        
        \ref{fig:invivo_mrilab}d) \& h) & \makecell{Hand / \\i3M} & 3D-RARE & $180\times180\times50$ & $120\times120\times10$ & 30 & 400 & 20 & \makecell{Center-out \\ ETL\,=\,10 \\ ESP\,=\,20\,ms} & 13 & 10.4 \\
        \hline
        
        \hline      
        \ref{fig:invivo_mrilab2}a) \& d) & \makecell{Calf / \\i3M} & 3D-RARE & $140\times140\times180$ & $90\times90\times36$ & 22.5 & 200 & 20 & \makecell{Center-out \\ ETL\,=\,5 \\ ESP\,=\,20\,ms} & 6 & 13 \\
        \hline
       
        \ref{fig:invivo_mrilab2}b) \& e) & \makecell{Calf / \\i3M} & 3D-RARE & $140\times140\times180$ & $90\times90\times18$ & 22.5 & 1000 & 60 & \makecell{Out-out \\ ETL\,=\,5 \\ ESP\,=\,20\,ms} & 2 & 11 \\
        \hline
        
        \ref{fig:invivo_mrilab2}c) \& f) & \makecell{Calf / \\i3M} & 3D-RARE & $120\times100\times180$ & $80\times60\times18$ & 20 & 1000 & 20 & \makecell{Center-out \\ ETL\,=\,5 \\ ESP\,=\,20\,ms} & 4 & 14 \\
        \hline
        
		\hline
	\end{tabular}
	\label{tab:seq_params}
\end{table*}

\section{Results \& Discussion}
\label{sec:results}

\subsection{Phantom images}
\label{sec:phantoms}

In the following sections we present phantom images acquired in the different setups: the i3M images have been taken with a variety of pulse sequences, the LUMC images show multi-Rx capabilities as required for parallel imaging, and the MGH section demonstrates the compatibility between \m{} and PulSeq.

\subsubsection{i3M}
\begin{figure}
	\centering
	\includegraphics[width=1\columnwidth]{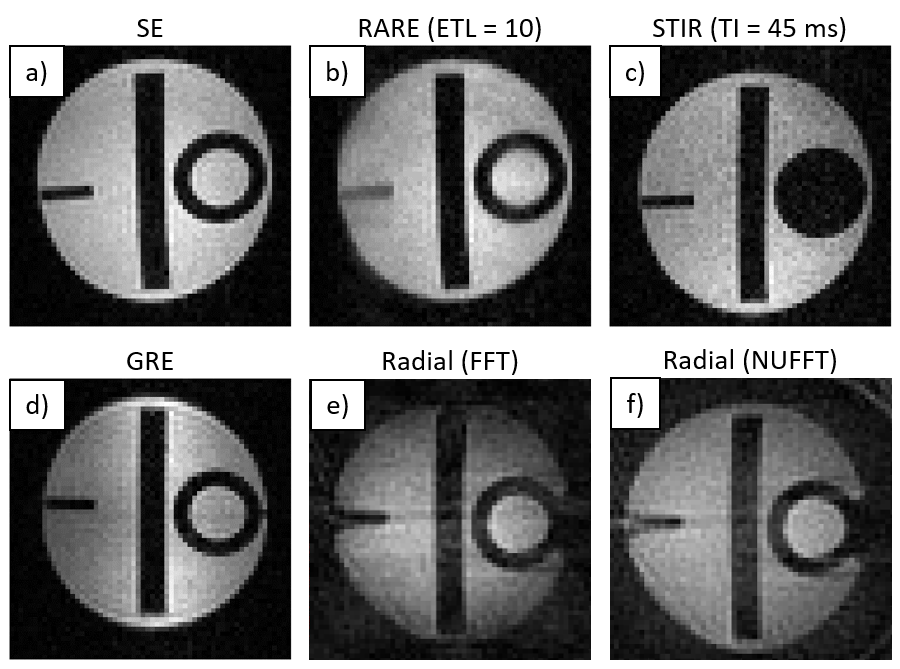}
	\caption{Images of a certified phantom \cite{Sipila2013}, acquired with different pulse sequences at i3M.}
	\label{fig:phantom_mrilab}
\end{figure}

First we show the performance of \m{} with five 3D pulse sequences and two reconstruction algorithms. To this end, we use a structured phantom used in the MRI accreditation program of the American College of Radiology \cite{Sipila2013}. The images in Fig.\,\ref{fig:phantom_mrilab} are all raw reconstructions, without filtering or post-processing. All pulse sequence parameters are included in Table\,\ref{tab:seq_params}. SE and GRE sequences (Fig.\,\ref{fig:phantom_mrilab}a) and d) respectively) are the simpler ones and serve as reference to evaluate the reconstruction quality of the rest. SE performs slightly better than GRE, since the former is less sensitive to magnetic field inhomogeneities. The RARE acquisition (Fig.\,\ref{fig:phantom_mrilab}b)) is run with an echo train length (ETL) of ten $\pi$-pulses. This shortens the scan time from 15\,min (Fig.\,\ref{fig:phantom_mrilab}a)) to 90\,s, but the image is slightly blurred due to small variable delays in the echo formation. The STIR sequence (Fig.\,\ref{fig:phantom_mrilab}c)) is executed with an inversion time (TI) of 45\,ms, previously calibrated to null the contribution from the smaller cylinder at the right in the phantom, which contains a substance with $T_1$ and $T_2$ characteristic times different from the rest of the sample. The difference in $T_1$ values is small, however, so this sequence retrieves significantly less signal per unit time than the rest. For the SNR in the image, the total scan time took around an hour. Finally, the images in Fig.\,\ref{fig:phantom_mrilab}e) and f) show reconstructions after a radial $k$-space acquisition. Both are reconstructed from the exact same data, one with a standard Fast Fourier Transform (FFT) after regridding with triangulation-based linear interpolation, the other with a NUFFT, where $k$-space points are weighted by density \cite{Greengard2004}. The latter performs notably better in terms of border definition, but the reconstruction is overall noisier and contains a circular artifact towards the outer region of the field of view. All in all, different sequences perform significantly different and can be advantageous for different applications. We find 3D-RARE most reliable in a large variety of situations, and have chosen it for the \emph{in vivo} acquisitions below.

\subsubsection{LUMC}
\begin{figure}
	\centering
	\includegraphics[width=0.75\columnwidth]{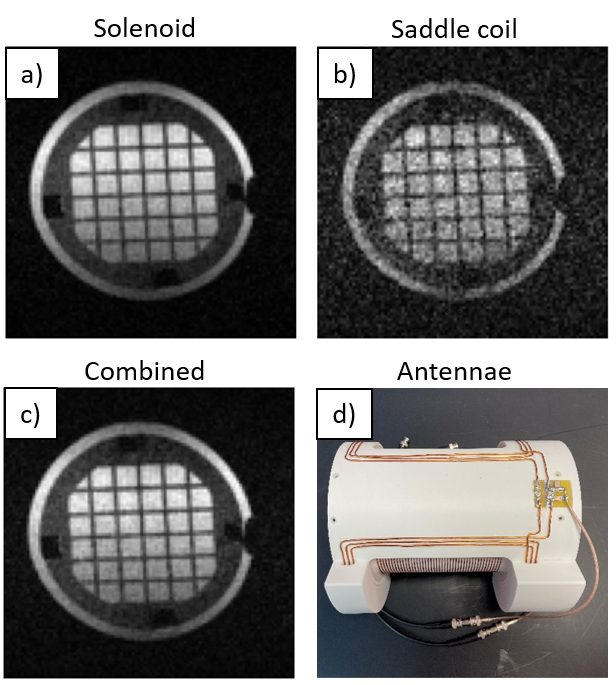}
	\caption{Images of a structured phantom acquired with simultaneous reception from two Rx coils at LUMC. a) Reconstruction using the data from the solenoid coil. b) Reconstruction using the data from the saddle coil. c) Reconstruction using the combined data from both coils. d) Photograph of the double-coil antenna, with the solenoid inside and the saddle coil outside.}
	\label{fig:phantom_lumc}
\end{figure}

In this experiment we evaluate the multi-Rx capabilities of \m{}. To this end we built a dual antenna comprising a saddle coil (188\,mm diameter, 202\,mm long) for both transmit and receive, and a solenoid (156\,mm diameter, 150\,mm long) for receive only (Fig.\,\ref{fig:phantom_lumc}d)). Coupling between the coils is inherently suppressed due to their orthogonal $B_1$ orientations ($S_{12} < 25$\,dB). The phantom (120\,mm diameter, 144\,mm long) images in Fig.\,\ref{fig:phantom_lumc} were acquired with a 3D-SE sequence (Table\,\ref{tab:seq_params}). The image data from the individual channels were filtered with a scaled sine-bell-squared filter prior to reconstruction with FFT. To combine the images, the data of each channel were normalized to their respective noise measurement prior to summing the magnitude images from the individual channels. The brightness of the single-Rx images was scaled from zero to the maximum value of each of the channels.

\subsubsection{MGH} 
\begin{figure}
	\centering
	\includegraphics[width=0.75\columnwidth]{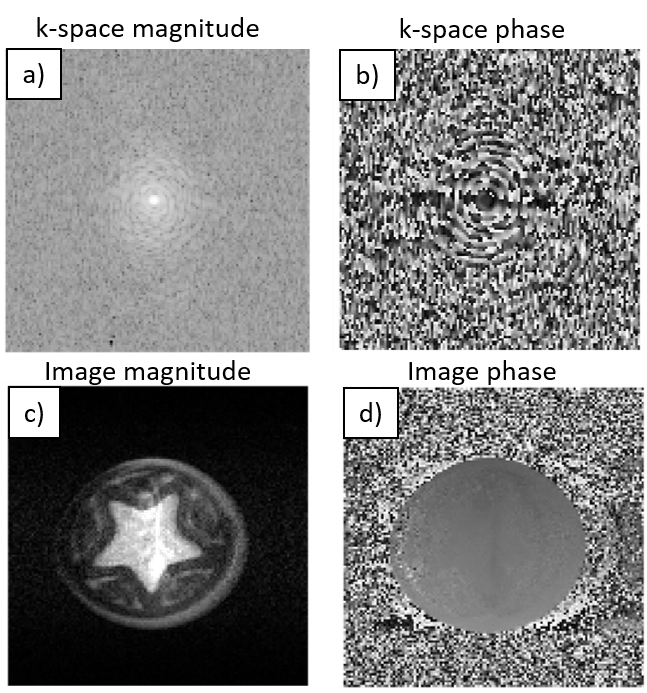}
	\caption{Image data of a star-shaped phantom acquired at MGH with a RARE spin-echo-train pulse sequence diagram generated by the open-source PulSeq sequence-programming environment in MATLAB. All excitation and refocusing RF pulses are \SI{80}{\micro s} long. a) $k$-space magnitude. b) $k$-space phase. c) Image magnitude. d) Image phase.}
	\label{fig:phantom_mgh}
\end{figure}

A Python interpreter was created to interface between the open-source PulSeq sequence programming environment and MaRCoS \cite{Layton2017}. Previous PulSeq interpreters were used with commercial high-field MRI scanners, but our work is the first extension to an open-source console. PulSeq uses simple syntax in Matlab \cite{PulSeq} or Python \cite{PyPulSeq} to generate a .seq text file that specifies all gradient, RF, and readout events occurring during a pulse sequence. This enables students and MRI physicists to rapidly code up sequences in the user-friendly PulSeq programming language. The .seq file is converted to console-specific executable code using the interpreter software. The PulSeq-MaRCoS interpreter software is available on GitHub \cite{MGHMARCOS}. Figure\,\ref{fig:phantom_mgh} shows image data acquired using the PulSeq-MaRCoS framework with a 2D-RARE spin-echo-train pulse sequence. The axial projection of the star-shaped phantom (10\,mm in diameter) is reconstructed from the acquired $k$-space data with FFT.

\subsection{\emph{In vivo} performance}
\label{sec:invivo}

\begin{figure*}
	\centering
	\includegraphics[width=1.5\columnwidth]{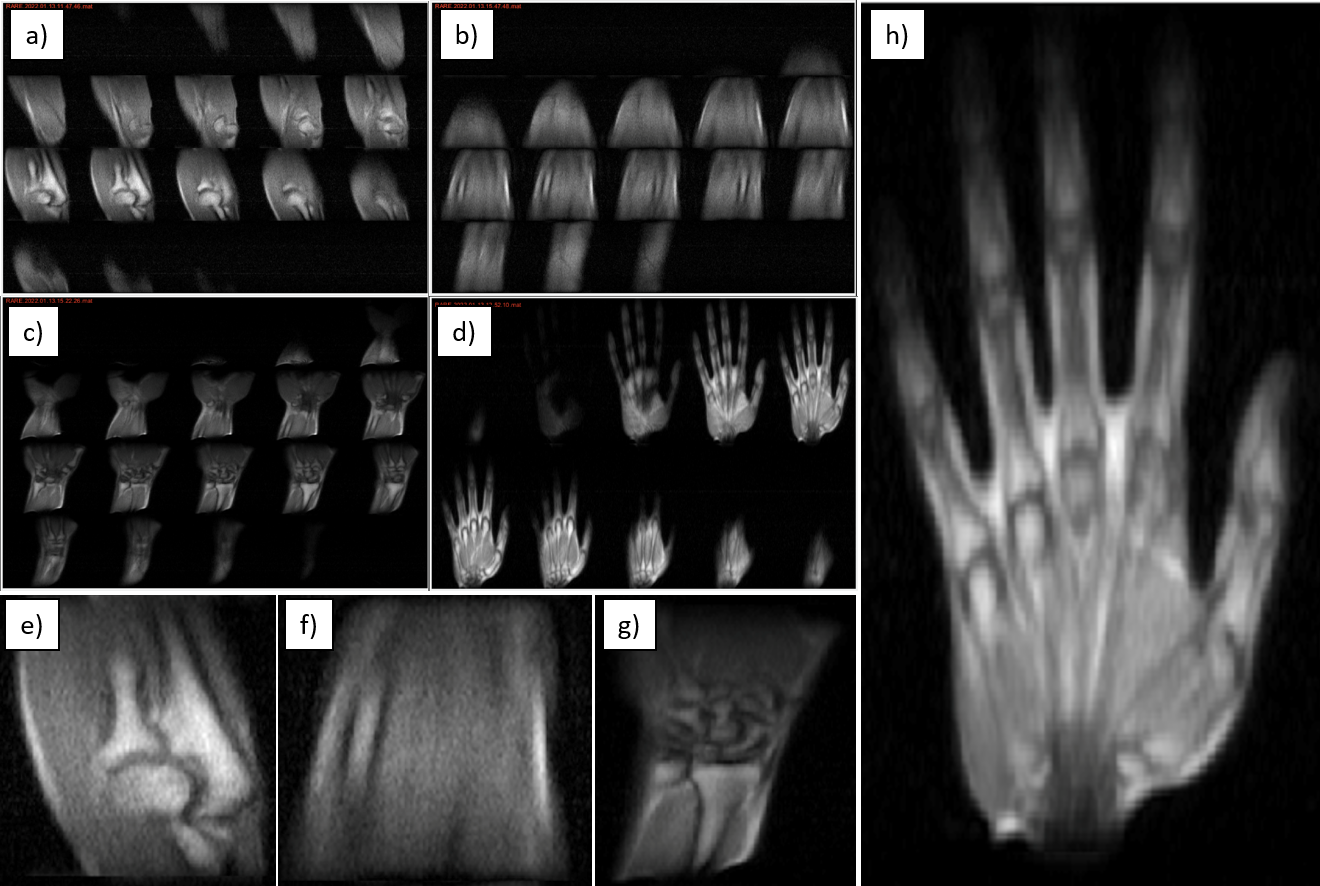}
	\caption{\emph{In vivo} images of the right upper limb taken at i3M with 3D-RARE sequences and reconstructed with FFT. Images a)-d) show all slices for the elbow, forearm, wrist and hand, respectively. Images e)-h) show magnified views of slices selected from a)-d).}
	\label{fig:invivo_mrilab}
\end{figure*}

The i3M \emph{in vivo} acquisitions in Fig.\,\ref{fig:invivo_mrilab} show images of the right upper limb of a healthy volunteer. These correspond to four independent scans of the hand, wrist, forearm and elbow. They are all $T_1$-weighted 3D-RARE acquisitions (see Table\,\ref{tab:seq_params}) and have been taken in less than 12\,min each. The SNR, contrast and spatial resolution suffice to identify different tissues in the anatomy, including muscle, fat, tendons, bone and bone marrow.

\begin{figure}
	\centering
	\includegraphics[width=1\columnwidth]{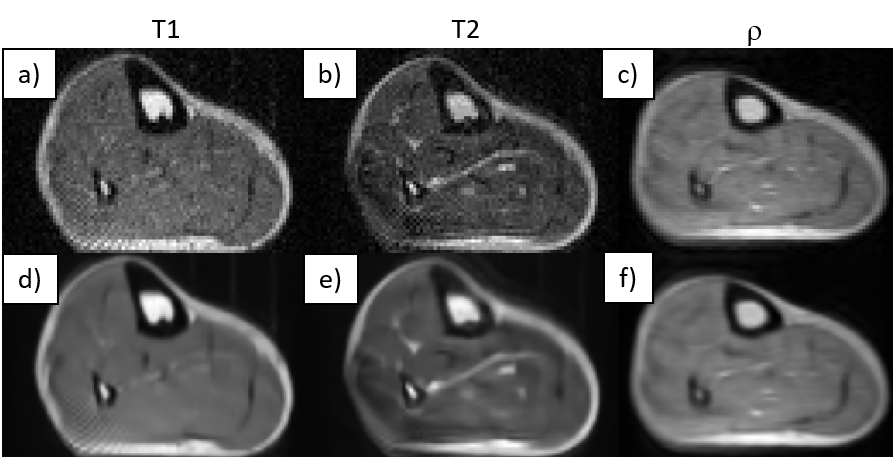}
	\caption{\emph{In vivo} images of the right calves of two healthy volunteers taken at i3M with 3D-RARE sequences and reconstructed with FFT. Images a)-c) show, respectively, a selected slices for the $T_1$, $T_2$ and $\rho$-weighted acquisitions. Images d)-f) are the same as a)-c), but BM4D-filtered. $T_1$ and $T_2$-weighted images are from the same subject.}
	\label{fig:invivo_mrilab2}
\end{figure}

To demonstrate different tissue contrasts, we acquired images from the right calves of two additional volunteers. The images in the top row of Fig.\,\ref{fig:invivo_mrilab2} correspond to $T_1$, $T_2$ and $\rho$-weighted acquisitions, where the first two are with the same subject. The bottom row shows the result of applying a BM4D filter \cite{Maggioni2013} to the above images. All pulse sequence parameters are provided in Table\,\ref{tab:seq_params}.

\section{Conclusion \& Outlook}
\label{sec:concl}

With this work we have demonstrated some of the main capabilities of low-field systems equipped with the first stable release of \m{}, including pulse sequences for \emph{in vivo} extremity imaging in clinically viable times and with sufficient quality for initial screening for common pathologies. Our results show that this platform is capable of driving many different system configurations, which has been possible thanks to collaboration and a commitment to detailed documentation. The different sequence programming capabilities, from the self-developed GUI to more widespread ``overlays'' such as PulSeq, greatly facilitates the development of pulse sequences to users with different demands and expertise. \m{} is in continuous evolution by virtue of an open, active, international network of collaborators, and we have planned a sizable upgrade for the coming months. This will include; new pulse sequences, such as steady-state free precession \cite{Sarracanie2015}, zero echo time \cite{Grodzki2012} and spiral \cite{Campbell2019}; data oversampling capabilities and advanced image reconstruction methods \cite{Galve2020b}; daisy-chaining of multiple Red Pitaya units for multi-Rx MRI beyond what we show in this work, as required for parallel imaging and acceleration techniques \cite{Pruessmann1999}; real-time decision-making capabilities during sequence execution, for potential optimization of pulses/gradients/delays via feedback control during data acquisition; and integration with post-processing and deep learning modules for sequence optimization, image reconstruction, denoising or segmentation tasks \cite{Lundervold2019}. The \m{} GUI will also continue to be developed to facilitate operation by users with different degrees of expertise, and to become readily usable in clinical environments by standardizing data formats and structure and complying with clinical procedural requirements. The \m{} community has the commitment to keep this project open-source and welcomes the participation of new users and contributors.

\appendices

\section*{Contributions}
TGN and JMA acquired data on the i3M scanner. RPG, TGN, JMA, FG and JA conceived and built the i3M scanner.
TOR acquired data on the LUMC scanner. TOR and AW conceived and built the LUMC scanner.
LCB wrote the Python interpreter to convert PulSeq files to MaRCoS/flocra format. JPS wrote PulSeq sequence code and acquired data on the MGH scanner.
VN and BM developed \m{}. YVG developed the \m{} GUI.
JA, AW and JPS wrote the paper, with input from all authors.

\section*{Acknowledgment}
We thank Thomas Witzel and Marcus Prier for discussions. This work was supported by the Ministerio de Ciencia e Innovaci\'on of Spain through research grant PID2019-111436RB-C21. Action co-financed by the European Union through the Programa Operativo del Fondo Europeo de Desarrollo Regional (FEDER) of the Comunitat Valenciana (IDIFEDER/2018/022 and IDIFEDER/2021/004) and Future and Emerging Technologies (FET, grant 101034644).

\section*{Ethical statement}
\emph{In vivo} experiments were carried out following Spanish regulations and under the research agreement from La Fe Hospital in Valencia (IIS-F-PG-22-02, agreement number 2019-139-1). Informed consents to participate and for publication were obtained from the volunteers prior to study commencement.

\section*{Code availability}
MaRCoS, the GUI and the PulSeq interface are publicly available from open-source repositories at \url{https://github.com/vnegnev/marcos_server}, \url{https://github.com/yvives/PhysioMRI_GUI} and \url{https://github.com/stockmann-lab/mgh_marcos}, respectively.

\section*{Conflict of interest}
TG is a researcher at Tesoro Imaging S.L. JMA, FG and JA are co-founders of PhysioMRI Tech S.L. JPS is a consultant for Neuro42, Inc.  All other authors declare no competing interests. 

\ifCLASSOPTIONcaptionsoff
  \newpage
\fi

\input{main.bbl}

\end{document}